\documentclass[prd,twocolumn,preprintnumbers,superscriptaddress,nofootinbib]{revtex4}
\usepackage[a4paper, hdivide={1.91cm,,1.165cm}, vdivide={1.83cm,,3.6cm}]{geometry}

\usepackage{amstext,amssymb}
\usepackage{amsmath}
\usepackage{graphicx}
\usepackage[hyperfootnotes=false]{hyperref}
\usepackage{xspace}
\usepackage{color}
\usepackage{units}
\usepackage{slashed} 

\begin{document}

\title{Implications of the diphoton excess on Left-Right models and gauge unification}
\author{Frank F. \surname{Deppisch}}
\email{f.deppisch@ucl.ac.uk}
\affiliation{Department of Physics and Astronomy, University College London, London WC1E 6BT, United Kingdom}
\author{Chandan \surname{Hati}}
\email{chandan@prl.res.in}
\affiliation{Physical Research Laboratory, Navrangpura, Ahmedabad 380 009, India}
\affiliation{Indian Institute of Technology Gandhinagar, Chandkheda, Ahmedabad 382 424, India}
\author{Sudhanwa \surname{Patra}}
\email{sudha.astro@gmail.com}
\affiliation{Center of Excellence in Theoretical and Mathematical Sciences,
Siksha 'O' Anusandhan University, Bhubaneswar-751030, India}
\author{Prativa \surname{Pritimita}}
\email{pratibha.pritimita@gmail.com}
\affiliation{Center of Excellence in Theoretical and Mathematical Sciences,
Siksha 'O' Anusandhan University, Bhubaneswar-751030, India}
\author{Utpal \surname{Sarkar}}
\email{utpal@prl.res.in}
\affiliation{Physical Research Laboratory, Navrangpura, Ahmedabad 380 009, India}

\begin{abstract}
The recent diphoton excess signal at an invariant mass of 750~GeV can be interpreted in the framework of left-right symmetric models with additional scalar singlets and vector-like fermions. We propose a minimal scenario for such a purpose. Extending the LRSM framework to include these new vector-like fermionic fields, on the other hand, results in interesting phenomenological implications for the LRSM fermion masses and mixing. Furthermore, existence of such vector-like fermions can also have interesting implications for baryogenesis and the dark matter sector. The introduction of a real bi-triplet scalar which contains a potential DM candidate will allow the gauge couplings to unify at $\approx 10^{17.7}$~GeV.
\end{abstract}

\maketitle 

\section{Introduction} 
The CMS and ATLAS collaborations have recently reported a roughly $3\sigma$ excess in the diphoton channel at an invariant mass of about 750~GeV in the first 3~${\rm{fb}}^{-1}$~of collected data from Run 2 of the LHC at 13~TeV \cite{CMS:2015dxe, ATLAS:Kado2015}. The Landau-Yang theorem forbids the possibility of a massive spin one resonance decaying to $\gamma\gamma$. The leading interpretations of the excess within the context of new physics scenarios therefore consist of postulating a fundamental spin zero or spin two particle with mass of about 750~GeV. However no enhancements have been seen in the dijet, $t \bar{t}$, diboson or dilepton channels posing a clear challenge to the possible interpretations of this excess. The absence of a peaked $\gamma\gamma$ angular distribution in the observed events towards the beam direction disfavours \cite{Ellis:2012wg} the spin two hypothesis and the spin zero resonance interpretation seems more favourable from a theoretical point of view.

A large number of interpretations of the diphoton signal in terms of physics beyond the Standard Model have been proposed in the literature \cite{DiChiara:2015vdm, Pilaftsis:2015ycr, Knapen:2015dap, Backovic:2015fnp, Molinaro:2015cwg, Gupta:2015zzs, Ellis:2015oso, Higaki:2015jag, Mambrini:2015wyu, Buttazzo:2015txu, Franceschini:2015kwy, Angelescu:2015uiz, Bellazzini:2015nxw, McDermott:2015sck, Low:2015qep, Petersson:2015mkr, Cao:2015pto, Kobakhidze:2015ldh, Agrawal:2015dbf, Chao:2015ttq, Fichet:2015vvy, Curtin:2015jcv, Csaki:2015vek, Aloni:2015mxa, Demidov:2015zqn, No:2015bsn, Bai:2015nbs, Matsuzaki:2015che, Dutta:2015wqh, Becirevic:2015fmu, Cox:2015ckc, Martinez:2015kmn, Bian:2015kjt, Chakrabortty:2015hff, Ahmed:2015uqt, Falkowski:2015swt, Cao:2009ah, Low:2012rj,Chao:2015nsm,Carpenter:2015ucu,Megias:2015ory,Alves:2015jgx,Gabrielli:2015dhk,Kim:2015ron,Benbrik:2015fyz,Pelaggi:2015knk,Hernandez:2015ywg,deBlas:2015hlv,Dev:2015isx,Boucenna:2015pav,Kulkarni:2015gzu,Chala:2015cev,Bauer:2015boy,Cline:2015msi,Berthier:2015vbb,Kim:2015ksf,Bi:2015uqd,Heckman:2015kqk,Huang:2015evq,Cao:2015twy,Wang:2015kuj,Antipin:2015kgh,Han:2015qqj,Ding:2015rxx,Chakraborty:2015jvs,Barducci:2015gtd,Cho:2015nxy,Feng:2015wil,Bardhan:2015hcr,Han:2015dlp,Dhuria:2015ufo,Chang:2015bzc,Han:2015cty,Arun:2015ubr,Li:2015jwd,Han:2015yjk,Casas:2015blx,Zhang:2015uuo,Liu:2015yec,Das:2015enc,Davoudiasl:2015cuo,Cvetic:2015vit,Badziak:2015zez,Patel:2015ulo,Moretti:2015pbj,Gu:2015lxj,Dev:2015vjd,Bizot:2015qqo,Anchordoqui:2015jxc,Goertz:2015nkp,Bi:2015lcf,Chao:2015nac,Gao:2015igz,Kim:2015xyn,Cao:2015scs,Cai:2015hzc,Wang:2015omi,Cao:2015apa,An:2015cgp,Tang:2015eko,Son:2015vfl,Park:2015ysf,Salvio:2015jgu,Chway:2015lzg,Ma:2015xmf,Kaneta:2015qpf,Hernandez:2015hrt,Low:2015qho,Dong:2015dxw,Kanemura:2015vcb,Kanemura:2015bli,Kang:2015roj,Chiang:2015tqz,Hamada:2015skp,Cao:2015xjz,Dasgupta:2015pbr,Chao:2016mtn,Murphy:2015kag,Chang:2015sdy,Huang:2015rkj,Marzola:2015xbh,Chakraborty:2015gyj,Nakai:2015ptz,Harigaya:2015ezk,Ghorbani:2016jdq,Hall:2015xds,Han:2016bus,Bonne:2012im,Angelescu:2015kga,Moreau:2012da,Allanach:2015ixl,Danielsson:2016nyy,Jiang:2015oms}. One of the possibilities that has been largely explored in the literature is a scalar or pseudo-scalar resonance produced through gluon-gluon fusion and decaying to $\gamma \gamma$ via loop diagrams with circulating fermions or bosons. A new resonance coupling with the Standard Model (SM) $t$ quark or $W^{\pm}$ can give rise to such loop diagrams, however, they will be highly suppressed at the large $\gamma \gamma$ invariant masses and the dominant decay channel would have to be $t \bar{t}$ or $W^{+}W^{-}$. Hence the observation of the $\gamma \gamma$ resonance at 750~GeV (much greater than the electroweak symmetry breaking scale) hints towards the existence of vector-like fermions around that mass scale. Given that both the ATLAS and CMS collaborations have suggested signal events consistent with each other at a tempting $3\sigma$ statistical significance level, hinting towards a new physics scenario, it is important to explore the possible model framework that can naturally accommodate such vector-like fermions.

From a theoretical stand point, a framework that can explain the diphoton excess while being consistent with other searches for new physics are particularly intriguing. To this end, one must mention the results reported by the CMS Collaboration in the first run of LHC for the right-handed gauge boson $W_R$ search at $\sqrt{s}= 8$~TeV and $19.7 {\rm{fb}}^{-1}$ of integrated luminosity \cite{Khachatryan:2014dka}. A $2.8\sigma$ local excess was reported in the $eejj$ channel in the energy range $1.8~\text{TeV} < m_{eejj} < 2.2$~TeV, hinting at a right handed gauge counterpart of the SM $SU(2)_{L}$ broken around the TeV scale. The Left-Right Symmetric Model (LRSM) framework with $g_R \neq g_L$ can explain such signal with the possibility of being embedded into a ultraviolet complete higher gauge group \cite{Deppisch:2014qpa, Deppisch:2014zta, Dev:2015pga,Deppisch:2015cua}. It is thus an interesting exercise to explore the possibility of naturally accommodating the $\gamma \gamma$ excess also in such a framework.

In this paper, we explore the possibility of extending the standard LRSM framework with vector-like fermions and singlet scalars which can explain the diphoton signal. Adding such new vector-like fermionic fields, on the other hand, results in interesting phenomenological implications for the LRSM fermion masses and mixing. Moreover, existence of such vector-like fermions can have interesting implications for baryogenesis and the potential dark matter sector. In gauged flavor groups with left-right symmetry \cite{Guadagnoli:2011id} or quark-lepton symmetric models \cite{Joshi:1991yn},  vector-like fermions are naturally accommodated while in LRSMs originating from D-brane or heterotic string compactifications also often include vector-like fermions \cite{Aldazabal:2000sk, Conlon:2008wa}. We propose a minimal LRSM that hosts such vector-like fermions and which can explain the diphoton signal. We also explore the possible fermion masses and mixing phenomenology and the implications of these vector-like particles in baryogenesis and the dark matter sector. 

The plan of rest of this paper is as follows. In section \ref{sec2}, we discuss the LRSM accommodating new vector-like fermions and the implications on masses and mixing of the fermions. In section \ref{sec3}, we discuss the general aspects of the diphoton signal in the context of a scalar resonance decaying to $\gamma\gamma$ through circulating vector-like fermions in the loop. In section \ref{unifn}, we explore the possibility of obtaining a gauge coupling unification including the new vector like fields. In section \ref{sec4}, we discuss the implications of the vector-like fermions for baryogenesis and the dark matter sector. In section \ref{sec5}, we summarize and make concluding remarks.


\section{Left-Right Symmetric Model Framework}{\label{sec2}}

The left-right symmetric extension of the SM has the basic gauge group given by
\begin{align}
	\mathcal{G}_{L,R}\equiv SU(3)_C \times SU(2)_{L} \times SU(2)_{R} \times U(1)_{B-L},
\end{align}
where $B-L$ is the difference between baryon and lepton number. The electric charge is related to the third component of isospin in the $SU(2)_{L,R}$ gauge groups and the $B-L$ charge as  
\begin{align}
	Q=T_{3L} + T_{3R} + 1/2(B-L).
\label{eq:charge}
\end{align}
The quarks and leptons transform under the LRSM gauge group as
\begin{align}
	q_{L} &= \begin{pmatrix} u_{L} \\ d_{L} \end{pmatrix}
	\equiv[2, 1, {\frac{1}{3}}, 3], \,\,
	q_{R} = \begin{pmatrix} u_{R}\\ d_{R}\end{pmatrix}
	\equiv[1,2,{\frac{1}{3}},3]\,,\nonumber \\
	\ell_{L} &= \begin{pmatrix}\nu_{L}\\ e_{L}\end{pmatrix}
	\equiv[2,1,-1,1], 
	\ell_{R} = \begin{pmatrix}\nu_{R}\\ e_{R}\end{pmatrix}\equiv[1,2,-1,1], \nonumber
\end{align}
where the gauge group representations are written in the form [$SU(2)_L, SU(2)_R, B-L, SU(3)_C$].

Originally the left-right symmetric extension of the SM \cite{Pati:1974yy, Mohapatra:1974gc, Senjanovic:1975rk, Senjanovic:1978ev, Mohapatra:1979ia, Mohapatra:1980yp}, was introduced to give a natural explanation for parity violation seen in radioactive beta decay and to consistently address the light neutrino masses via seesaw mechanism \cite{Minkowski:1977sc, GellMann:1980vs, Yanagida:1979as, Schechter:1980gr, Schechter:1981cv}. The latter form doublets with the right handed charged fermions under the $SU(2)_R$ gauge group. If $SU(2)_R$ breaks at around the TeV scale, LRSMs offer a rich interplay between high energy collider signals and low energy processes such as neutrinoless double beta decay and lepton flavor violation \cite{Patra:2012ur}. The principal prediction of this scenario is a TeV scale right-handed gauge boson $W_R$. The CMS and ATLAS collaborations had reported several excesses at around 2~TeV in Run 1 of the LHC, pointing towards such a possibility. From the first results of Run 2,  no dijet and diboson excesses have been reported (more data is required to exclude the diboson excesses reported in run 1), the relatively ``cleaner" $eejj$ channel signal is still present hinting at a 2~TeV $W_R$. In light of the diphoton excess it is important to revisit the LRSM framework to explore the possibility of accommodating such signal and the possible implications. 

As already mentioned in the introduction and discussed in section \ref{sec3}, the 750~GeV  diphoton excess can be explained through the resonant production and decay of a scalar or pseudoscalar particle. To this end, we propose a simple left-right symmetric model with a scalar singlet $S$ and vector-like fermions added to the minimal particle content of left-right symmetric models \footnote{We assume the resonance to be a new singlet scalar and it can easily be generalized to a pseudoscalar case.}.  

We extend the standard LRSM framework to include isosinglet vector-like copies of LRSM  fermions. This kind of a vector-like fermion spectrum is very naturally embedded in gauged flavour groups with left-right symmetry \cite{Guadagnoli:2011id} or quark-lepton symmetric models \cite{Joshi:1991yn}. The field content of this model and the relevant transformations under the LRSM gauge group are shown in Tab.~\ref{tab:LR2}.
\begin{table}[h]
\begin{center}
\begin{tabular}{|c|c|c||c|c|}
\hline
Field     & $ SU(2)_L$ & $SU(2)_R$ & $B-L$ & $SU(3)_C$ \\
\hline
$q_L$     &  2         & 1         & 1/3   & 3   \\
$q_R$     &  1         & 2         & 1/3   & 3   \\
$\ell_L$  &  2         & 1         & -1    & 1   \\
$\ell_R$  &  1         & 2         & -1    & 1   \\
\hline
$U_{L,R}$ &  1         & 1         & 4/3   & 3   \\
$D_{L,R}$ &  1         & 1         & -2/3  & 3   \\
$E_{L,R}$ &  1         & 1         & -2    & 1   \\
$N_{L,R}$ &  1         & 1         & 0     & 1   \\
\hline
 $H_L$    &  2         & 1         & 1    & 1   \\
 $H_R$    &  1         & 2         & 1    & 1   \\
 $S$      &  1         & 1         & 0     & 1   \\
\hline
\end{tabular}
\end{center}
\caption{LRSM representations of extended field content.}
\label{tab:LR2}
\end{table}

The relevant Yukawa part of the Lagrangian is given by
\begin{align}
\mathcal{L} = 
	&- \sum_{X} ( \lambda_{SXX} S \overline{X} X + M_X \overline{X} X ) \nonumber\\
	&- (\lambda_U^L \tilde{H}_L \overline{q}_L U_R 
     + \lambda_U^R \tilde{H}_R \overline{q}_R U_L \nonumber\\
  &+\phantom{(}\lambda_D^L H_L \overline{q}_L D_R
	   + \lambda_D^R H_R \overline{q}_R D_L \nonumber\\
  &+\phantom{(}\lambda_E^L H_L\overline{\ell}_L E_R 
	   + \lambda_E^R H_R\overline{\ell}_R E_L \nonumber\\
  &+\phantom{(}\lambda_N^L \tilde{H}_L \overline{\ell}_L N_R 
	   + \lambda_N^R \tilde{H}_R \overline{\ell}_R N_L + \text{h.c.}),
\label{2.1}
\end{align}
where the summation is over $X = U, D, E, N$ and we suppress flavour and colour indices on the fields and couplings. $\tilde{H}_{L,R}$ denotes $\tau_2 H_{L,R}^\ast$, where $\tau_2$ is the usual second Pauli matrix.

The vacuum expectation values (VEVs) of the Higgs doublets  $H_R(1,2,-1)$ and $H_L(2,1,-1)$ break the LRSM gauge group to the SM gauge group and the SM gauge group to $U(1)_\text{EM}$ respectively, with an ambiguity of parity breaking, which can either be broken at the TeV scale or at a much higher scale $M_P$. In the latter case, the Yukawa couplings can be different for right-type and left-type Yukawa terms because of the renormalization group running below $M_P$, $\lambda_X^R \neq \lambda_X^L$. Hence, we distinguish the left and right handed couplings explicitly with the subscripts $L$ and $R$. We use the VEV normalizations $\langle H_{L} \rangle= ( 0, v_L)^T$ and $\langle H_{R} \rangle = (0, v_R)^T$ with $v_L = 175$~GeV and $v_R$ constrained by searches for the heavy right-handed $W_R$ boson at colliders and at low energies, $v_R \gtrsim 1 - 3$~TeV (depending on the right-handed gauge coupling).
Due to the absence of a bidoublet Higgs scalar, normal Dirac mass terms for the SM fermions are absent and the charged fermion mass matrices assume a seesaw structure. However, if one does not want to depend on a ``universal" seesaw structure, a Higgs bidoublet $\Phi$ can be introduced along with $H_{L,R}$.

After symmetry breaking, the mass matrices for the fermions are given by
\begin{align}
\label{2.3}
	M_{uU}    = \begin{pmatrix} 0 & \lambda_U^L v_L \\ \lambda_U^R v_R & M_U \end{pmatrix}, \,
	M_{dD}    = \begin{pmatrix} 0 & \lambda_D^L v_L \\ \lambda_D^R v_R & M_D \end{pmatrix}, 
	\nonumber\\
	M_{e E}   = \begin{pmatrix} 0 & \lambda_E^L v_L \\ \lambda_E^R v_R & M_E \end{pmatrix}, \,
	M_{\nu N} = \begin{pmatrix} 0 & \lambda_N^L v_L \\ \lambda_N^R v_R & M_N \end{pmatrix},
\end{align}
The mass eigenstates can be found by rotating the mass matrices via left and right orthogonal transformations $O^{L,R}$ (we assume all parameters to be real). For example, the up quark diagonalization yields $O^{LT}_U \cdot M_{uU} \cdot O^R_U = \text{diag}(\hat{m}_u, \hat{M}_U)$. Up to leading order in $\lambda^L_U v_L$, the resulting up-quark masses are
\begin{align}
\label{2.4.0}
	\hat{M}_U \approx \sqrt{M_U^2 + (\lambda^R_U v_R)^2}, \quad
	\hat{m}_u \approx \frac{(\lambda^L_U v_L)(\lambda^R_U v_R)}{\hat{M}_U},
\end{align}
and the mixing angles $\theta^{L,R}_U$ parametrizing $O^{L,R}_U$,
\begin{align}
\label{2.4.1}
	\tan(2\theta^L_U) \approx \frac{2(\lambda^L_U v_L) M_U}{M_U^2 + (\lambda^R_U v_R)^2},
	\tan(2\theta^R_U) \approx \frac{2(\lambda^R_U v_R) M_U}{M_U^2 - (\lambda^R_U v_R)^2}.
\end{align}
The other fermion masses and mixings are given analogously. For an order of magnitude estimate one may approximate the phenomenologically interesting regime with the limit $\lambda^R_U v_R \to M_U$ in which case the mixing angles approach $\theta^L_U \to \hat{m}_u / \hat{M}_U$ and $\theta^R_U \to \pi/4$. This means that $\theta^L_U$ is negligible for all fermions but the top quark and its vector partner \cite{Dev:2015vjd}.

We here neglect the flavour structure of the Yukawa couplings $\lambda^{L,R}_X$ and $\lambda_{SXX}$ which will determine the observed quark and leptonic mixing. The hierarchy of SM fermion masses can be generated by either a hierarchy in the Yukawa couplings or in the masses of the of the vector like fermions.

As described above, the light neutrino masses are of Dirac-type as well, analogously given by
\begin{equation}
\label{2.7}
	\hat{m}_\nu = \frac{\lambda^L_N \lambda^R_N v_L v_R}{M_N},
\end{equation}
It is natural to assume that $M_N \gg v_R$, as the vector like $N$ is a singlet under the model gauge group. In this case, the scenario predicts naturally light Dirac neutrinos \cite{Guadagnoli:2011id}.

\section{Diphoton signal from a scalar resonance}{\label{sec3}}
One may attempt to interpret the diphoton excess at as the resonant production of the singlet scalar $S$ with mass $M_S = 750$~GeV. Considering the possible production mechanisms for the resonance at 750~GeV it is interesting to note that the CMS and ATLAS did not report a signal in the $\sim 20 {\rm{fb}}^{-1}$ data at 8~TeV in Run 1. One possible interpretation of this can be that the resonance at 750~GeV is produced through a mechanism with a steeper energy dependence. Excluding the possibility of an associated production of this resonance, the most favourable mechanism is gluon-gluon fusion which we here also consider as the dominant production mechanism. Subsequently, the scalar with mass 750~GeV decays to two photons via a loop as well. The cross section can be expressed as
\begin{align}
	\label{eq:cs}
	\sigma(pp \to \gamma \gamma) = \frac{C_{gg}}{M_S s} 
	\Gamma_{gg} \text{Br}_{\gamma\gamma},
\end{align}
with the proton centre of mass energy $\sqrt{s}$ and the parton distribution integral $C_{gg} = 174$ at $\sqrt{s} = 8$~TeV and $C_{gg} = 2137$ at $\sqrt{s} = 13$~TeV \cite{Franceschini:2015kwy}. One can obtain a best fit guess of the cross section by reconstructing the likelihood, assumed to be Gaussian, from the $95\%$ C.L. expected and observed limits in an experimental search. For the diphoton excess, we use a best fit cross section value of 7~fb found by combining the 95\% CL ranges from ATLAS and CMS at 13~TeV and 8~TeV for a resonance mass of 750~GeV \cite{Franceschini:2015kwy}.

Apart from the necessary decay modes of the scalar $S$ i.e, $S\to gg$ and $S\to \gamma\gamma$, $S$ may also decay to other particles; due to the necessary SM invariance and the fact that $M_S > m_Z$, $S\to \gamma\gamma$ necessitates the decays $S\to \gamma Z$ and $Z Z$ which are suppressed by $2\tan^2 \theta_W \approx 0.6$ and $\tan^4\theta_W \approx 0.1$ relative to $\Gamma(S \to \gamma\gamma)$ \cite{Franceschini:2015kwy}. Furthermore, $S$ in our model may also decay to SM fermions due to mixing with the heavy vector-like fermions. As described above, the mixing is only sizeable for the top and its vector partner. The total width is thereby given by $\Gamma_S \approx \Gamma_{gg} + 1.7\times\Gamma_{\gamma\gamma} + \Gamma_{t\bar{t}}$.

We would like to stress that while the diphoton resonance undoubtedly is the motivation behind this work, the purpose of this paper is to construct a consistent LRSM framework that can naturally accommodate vector-like fermions taking the diphoton signal as a hint and explore the consequent phenomenology. One can find similar interpretations of the diphoton excess in \cite{Cao:2015xjz, Dev:2015vjd, Dasgupta:2015pbr} in models with a singlet scalar accompanied by vector-like fermions.

Production of a scalar resonance in gluon fusion via a loop of vector-like quarks and subsequent 
decay of scalar resonance to $\gamma\gamma$ via a loop of vector-like quarks and leptons. There are contributions to $\Gamma(S\to \gamma \gamma)$ from quark-like vector fermion $\psi_Q=U,D$ and lepton-like vector fermion $\psi_L=E$ propagating inside the loop. Apart from quark-like vector fermion contributing to the production of scalar through gluon fusion, there could be another top-quark mediated diagram via mixing with SM Higgs boson.

In the LRSM framework discussed in section \ref{sec2}, the vector-like degrees of freedom contribute to the loop leading to $S\to gg$ and $S\to \gamma\gamma$. The partial decay widths are given by \cite{McDermott:2015sck}
\begin{align}
\label{3.8}
	\Gamma_{\gamma\gamma} &= \phantom{K}
		\frac{\alpha^2\, M^3_S}{256 \pi^3} 
    \left| \sum_X
			\frac{N^C_X Q^2_X \lambda'_{SXX}}{M_X} 
			\mathcal{A}\left(\frac{m_S^2}{4 M_X^2} \right) 
		\right|^2, \nonumber\\
	\Gamma_{gg} &=
		K \frac{\alpha^2_s M^3_S}{128 \pi^3} 
		\left| \sum_X^C
			\frac{\lambda'_{SXX}}{M_X} 
			\mathcal{A}\left(\frac{m_S^2}{4 M_X^2} \right)
		\right|^2.
\end{align}
Here, the sums in $\Gamma_{\gamma\gamma}$ and $\Gamma_{gg}$ are over all and coloured fermion species and flavours, respectively. $N_X^C$ is the number of color degrees of freedom of a species, i.e 1 for leptonic vector-like fermions and 3 for quark-like fermions. Similarly, $Q_X$ is the electric charge of the species. The effective coupling of $S$ to a fermion species is $\lambda'_{SXX} = \lambda_{SXX} (O^R_X)_{1X}(O^L_X)_{1X}$, i.e. the coupling $\lambda_{SXX}$ dressed with the corresponding left and right mixing matrix element. The value of the parameters $\alpha \approx 1/127$, $\alpha_s \approx 0.1$ and $K \approx 1.7$ \cite{McDermott:2015sck} in which $A(x)$ is a loop function defined by
\begin{align}
\label{3.9}
	A(x) = \frac{2}{x^2} [x + (x - 1) f(x)],
\end{align}
with
\begin{align}
\label{3.10}
	f(x) = \left\{ 
		\begin{matrix}
			\arcsin^{2} \sqrt{x} & & x \leq 1 \\ 
	    -\frac{1}{4}\left[\ln\left(\frac{1+\sqrt{1-x}}{1-\sqrt{1-x}}\right)-i\pi\right]^2 
			& & x > 1.
		\end{matrix} 
	\right.
\end{align}

In addition, the decay of $S$ to a pair of fermions (here only relevant for the top) is given by
\begin{align}
\label{3.11}
	\Gamma_{f\bar{f}} = \frac{N_f^C \lambda'^2_{Xff} M_S}{16\pi}
	\left(1 - \frac{4 M_f^2}{M_S^2} \right)^{2/3}.
\end{align}

In order to arrive at an estimate for the diphoton production cross section, we assume that the vector fermion masses and couplings to $S$ are degenerate ($M_X$, $\lambda_{SXX}$), except for the the top partner ($M_T$, $\lambda_{STT}$). In the limit of large vector fermion masses $M_X \gtrsim M_S/2$, we arrive at the approximation for the partial widths,
\begin{align}
\label{3.12}
	\frac{\Gamma_{gg}}{M_S} &\approx 
		1.3 \times 10^{-4} \left(\frac{\lambda_{SXX} \cdot \text{TeV}}{M_X}\right)^2, \nonumber\\
	\frac{\Gamma_{\gamma\gamma}}{M_S} &\approx 
		3.4 \times 10^{-7}  \left(\frac{\lambda_{SXX} \text{TeV}}{M_X}\right)^2, \nonumber\\	
	\frac{\Gamma_{t\bar{t}}}{M_S} &\approx 
		1.3 \times 10^{-3} \left(\frac{\lambda_{TXX} \text{TeV}}{M_T}\right)^2.	
\end{align}
As discussed in \cite{Franceschini:2015kwy} in a model-independent fashion, the diphoton excess can be explained for $10^{-6} \lesssim \Gamma_{gg}/M_S \lesssim 2 \times 10^{-3}$ (the upper limit is due to the limit from dijet searches) and $\Gamma_{\gamma\gamma}/M_S \approx 10^{-6}$, as long as $gg$ and $\gamma\gamma$ are the only decay modes of $S$. In order to achieve this, the top partner $T$ needs to have a significantly weaker coupling or heavier mass than the rest of the vector fermions. Assuming the decay width to $t\bar{t}$ contributes negligibly to the total width, the diphoton cross section, Eq.~\eqref{eq:cs} is
\begin{align}
	\sigma(pp \to \gamma \gamma) \approx 1.7\text{ fb} \cdot
		\left(\frac{\lambda_{SXX} \cdot \text{TeV}}{M_X}\right)^2,
\end{align}
The experimentally suggested cross section $\sigma(pp \to \gamma \gamma) \approx 7$~fb can be achieved with $M_X / \lambda_{SXX} \approx 0.5$~TeV (and $\Gamma_{gg}/M_S \approx 5\times 10^{-4}$ satisfying the dijet limit). In such a scenario, the total width of $S$ is of the order $\Gamma_S \approx 0.5$~GeV, i.e. much smaller than the 45~GeV suggested by ATLAS if interpreted as a single particle resonance. $\Gamma_{\gamma\gamma}/\Gamma_{gg}$ can also be independently boosted by introducing a hierarchy with leptonic partners lighter than the quark partners. While certainly marginal and requiring a specific structure among the vector fermions, this demonstrates that the diphoton excess, apart from the broad width seen by ATLAS, can be accommodated in our model.


\section{Gauge Coupling Unification}{\label{unifn}} 
In the previous section, we have discussed how the inclusion of new vector-like fermions in LRSM can aptly explain the diphoton excess traced around 750 GeV at the LHC. Interestingly this framework can also be embedded in a non-SUSY grand unified theory like $SO(10)$ having left-right symmetry as its only intermediate symmetry breaking step with the breaking chain given as follows
\begin{align}
	SO(10) \mathop{\longrightarrow}^{\langle \Sigma \rangle} \mathcal{G}_{2213P}  
				 \mathop{\longrightarrow}^{\langle H_R    \rangle} \mathcal{G}_{213} 
				 \mathop{\longrightarrow}^{\langle H_{L}  \rangle} \mathcal{G}_{13}. 
\end{align}
The $SO(10)$ group breaks down to left-right symmetric group $\mathcal{G}_{2213P} \equiv SU(2)_L \times SU(2)_R \times U(1)_{B-L} \times SU(3) \times \mathcal{P}$ via a non-zero VEV of $\Sigma \subset 210_H$. Here, $\mathcal{P}$ is defined as the discrete left-right symmetry, a generalized parity symmetry or charge-conjugation symmetry. The vital step is to break the left-right gauge symmetry and this is attained with the help of the right-handed Higgs doublet $H_R$. Finally, the SM gauge group is spontaneously broken by its left-handed counterpart $H_L$. As described above we add another scalar singlet $S$ in order to explain the diphoton signal though it is not contributing to the renormalization group (RG) evolution of the gauge couplings. 

\begin{figure}[t]
\includegraphics[scale=0.65]{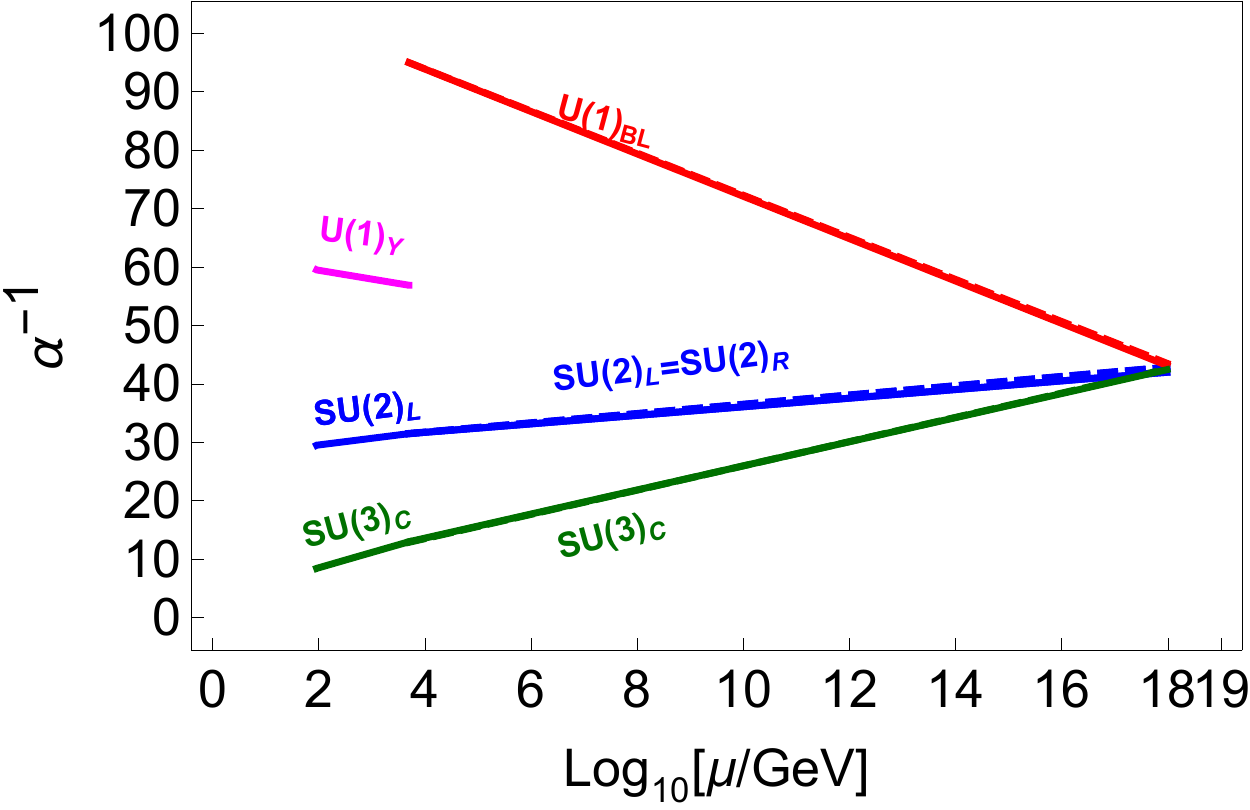}
\caption{Gauge coupling running in the considered model accommodating the diphoton excess, demonstrating successful gauge unification at the scale $M_\text{GUT} = 10^{17.75}$~GeV with an intermediate left-right symmetry breaking scale at 10~TeV. The dashed lines correspond to one loop RGE of gauge couplings while the two loop effects are displayed in solid lines.} 
\label{fig:unifn}
\end{figure}

In addition to the particle content described in Tab.~\ref{tab:LR2}, we include a bi-triplet $\eta \equiv (3,3,0,1)$ under $\mathcal{G}_{2213P}$ to achieve successful gauge unification. This can be confirmed by using the relevant RG equation for the gauge couplings $g_i$,
\begin{equation}
\mu\,\frac{\partial g_{i}}{\partial \mu}=\frac{b_i}{16 \pi^2} g^{3}_{i},
\end{equation}
where the one-loop beta-coefficients $b_i$ are given by
\begin{eqnarray}
	&&b_i= - \frac{11}{3} \mathcal{C}_{2}(G) 
				 + \frac{2}{3} \,\sum_{R_f} T(R_f) \prod_{j \neq i} d_j(R_f) \nonumber \\
  &&\hspace*{2.5cm} + \frac{1}{3} \sum_{R_s} T(R_s) \prod_{j \neq i} d_j(R_s).
\label{oneloop_bi}
\end{eqnarray}
Here, $\mathcal{C}_2(G)$ is the quadratic Casimir operator for gauge bosons in their adjoint representation,
\begin{equation}
	\mathcal{C}_2(G) \equiv 
	\begin{cases} 
		N & \text{if } SU(N), \\
    0 & \text{if }  U(1).
	\end{cases}
\end{equation}
$T(R_f)$ and $T(R_s)$ are the traces of the irreducible representation $R_{f,s}$ for a given fermion and scalar, respectively,
\begin{equation}
	T(R_{f,s}) \equiv 
	\begin{cases} 
		1/2 & \text{if } R_{f,s} \text{ is fundamental}, \\
    N   & \text{if } R_{f,s} \text{ is adjoint}, \\
		0   & \text{if } U(1).
	\end{cases}
\end{equation}
and $d(R_{f,s})$ is the dimension of a given representation $R_{f,s}$ under all $SU(N)$ gauge groups except the $i$-th~gauge group under consideration. An additional factor of $1/2$ should be multiplied in the case of a real Higgs representation. Using the above particle content, the beta-coefficients at one loop are found to be $b_{2L} = -19/6$, $b_{Y} = 41/10$, $b_{3C} = -7$ from the SM to the LR breaking scale and $b_{2L} = b_{2R} = -13/6$, $b_{BL}= 59/6$, $b_{3C} = -17/3$ from the LR breaking scale to the GUT scale. We have also evaluated the two loop contributions which give a very marginal deviation over one loop contributions. The resulting running of the gauge couplings at one loop and two loop orders are shown in Fig.~\ref{fig:unifn} with the breaking scales
\begin{align}
	M_\text{GUT} = 10^{17.75} \text{ GeV}, \quad M_\text{LR} = 10\text{ TeV}.
\end{align}
%

\section{Implications for baryogenesis and dark matter}{\label{sec4}} 

The vector-like fermions added to the spectrum of the LRSM framework can have very profound implications for a baryogenesis mechanism such as leptogenesis and the dark matter sector. While the proposal of high scale leptogenesis via singlet heavy Majorana neutrinos (or a heavy Higgs triplet) decay added to the SM is beyond the reach of the present and near future collider experiments, the LRSM scenario provides a window of opportunity for low TeV scale leptogenesis testable at the LHC. However, the observation of a 2~TeV $W_R$ boson at the LHC, through confirmation of the $2.8\sigma$ signal of two leptons and two jets reported by the CMS collaboration, would rule out the possibility of high scale as well as TeV scale resonant leptogenesis with the standard LRSM fields due to the unavoidable fast gauge mediated $B-L$ violating interactions \cite{Ma:1998sq, Frere:2008ct, Deppisch:2015yqa, Deppisch:2013jxa, Dev:2014iva, Dhuria:2015wwa, Dhuria:2015cfa, Dev:2015vra}. 

On the other hand, the new vector-like fermions added to the LRSM to accommodate the diphoton excess can open a whole new world of possibilities. A particularly interesting possibility is the realization of baryogenesis and dark matter annihilation through a vector-like portal first explored in \cite{Perez:2013nra}. As an example, consider the following additional terms in the Lagrangian,
\begin{align}{\label{4.1}}
	\mathcal{L} \supset &
		-(\lambda_{XU} X \overline{u_R} U_L + \text{h.c.}) 
		- m_X^2 X^\dagger X -\lambda_X (X^\dagger X)^2\nonumber\\
	& -\lambda_{HX} H^\dagger H X^\dagger X,
\end{align} 
where $X$ is an inert doublet (a singlet complex) dark matter scalar field in the LR(SM) case. $X$ is charged under some exotic global $U(1)_\chi$ symmetry, under which only the vector-like quarks and dark matter fields transform non-trivially. Thus, the introduction of a vector-like quarks can connect the dark matter to the usual LR(SM) quarks, which can be readily used to make a connection between the baryon asymmetry and dark matter, as pointed out in \cite{Perez:2013nra}. In rest of this section we focus on sketching the simpler case of the extended SM which can be expanded to the LRSM case by replacing the singlets with appropriate doublet representations. However, in the case of the LRSM some subtleties are present and we comment on them towards the end of this section. On the other hand, this idea can easily be generalized to accommodate a down-type quark portal or charged lepton portal (corresponding to a leptogenesis scenario of baryogenesis).

The basic idea behind the vector-like portal is to generate an asymmetry in the vector-like sector through baryogenesis, which then subsequently gets transferred to the SM baryons and the dark matter sector through the renormalizable couplings in Eq.~\eqref{4.1}. In addition one can introduce a scalar field $Y$ with the couplings
\begin{align}{\label{4.2}}
	\mathcal{L} \supset 
		& -\left(\lambda_{\nu Y} Y\nu_R \nu_R + \rm{h.c.}\right)
		  -m_Y^2 Y^\dagger Y -\lambda_Y(Y^\dagger Y)^2 \nonumber\\
	 &  -\lambda_{XY} X^\dagger X Y^\dagger Y,
\end{align} 
which allows the annihilation of a pair of $X$ into $Y$ fields. The latter can subsequently decay into two singlet right handed neutrinos ensuring the asymmetric nature of the dark matter $X$ relic density for a large enough annihilation cross section. Now turning to the question of how to generate the primordial asymmetry in the vector-like sector which defines the final dark matter asymmetry and baryon asymmetry, let us further add two types of heavy diquarks with the couplings
\begin{align}{\label{4.3}}
	\mathcal{L} &\supset 
		  \lambda_{\Delta U_L} \Delta_u U_L U_L 
		+ \lambda_{\Delta U_R} \Delta_u U_R U_R 
		+ \lambda_{\Delta d}   \Delta_d d_R d_R \nonumber\\
	 &+ \lambda_\chi \Delta_u \Delta_d \Delta_d \chi + \rm{h.c.},
\end{align} 
where $\Delta_{u}: (\bar{6},1,-4/3)$, $\Delta_{d}: (\bar{6},1,2/3)$ and the field $\chi$ breaks the local $U(1)_\chi$ symmetry under which $X$ and $U$ have non-trivial charges denoted by $q_\chi(U)$ and $q_\chi(X)$. For the SM fields this charge is simply $B-L$, which right away gives $q_\chi(\Delta_d) = -2/3$. The rest of the charges are determined in terms of the free charge $q_\chi(U)$,
\begin{align}{\label{4.4}}
	q_\chi(\Delta_u) &= -2q_\chi(U), q_\chi(\chi) = 2q_\chi(U) + 4/3,\nonumber\\
	q_\chi(X)        &= 1/3 - q_\chi(U).
\end{align}
In order too forbid the dangerous proton decay induced by the operators $\mathcal{O} = X^2, S^2, X^2 S^2, X^4, S^4$ \cite{Arnold:2012sd}, one needs to satisfy the condition
\begin{align}{\label{4.5}}
q_{\chi}(\mathcal{O})\neq n (2q_{\chi}(U)+4/3), \;\rm{where} \; n=0, \pm 1, \pm 2, \cdots .
\end{align}
From Eq.~\eqref{4.3} it follows that after $\chi$ acquires a VEV to break the $U(1)_\chi$ symmetry, $\Delta_u$ has the decay modes
\begin{align}{\label{4.6}}
	\Delta_u \to \Delta_d^\ast \Delta_d^\ast, \;\; \Delta_u \to \bar{U}\bar{U},
\end{align}
and a $CP$ asymmetry (between the above modes and their conjugate modes) can be obtained by interference of the tree level diagrams with one loop self energy diagrams with two generations of $\Delta_u$. Finally, the asymmetry generated in the vector-like quarks gets transferred to the dark matter asymmetry and baryon asymmetry via the $\lambda_{XU}$ term in Eq.~\eqref{4.1}. This mechanism gives a ratio between the dark matter relic density and the baryon asymmetry given by
\begin{align}{\label{4.7}}
	\frac{\Omega_{DM}/m_X}{\Omega_B/M_p} = \frac{79}{28},
\end{align}
and in this model a dark matter mass $m_{X}\sim 2$~GeV. A typical prediction of this model is neutron-antineutron oscillations induced by the up-type and two down-type diquarks through the mixing of vector-like up-type quarks with the usual up quarks. However, such oscillations will be suppressed by the mixing.

One can similarly construct a leptogenesis model involving vector-like charged leptons. In case of the LRSM a generalization of the above scheme is straightforward; however, the lepton number violating gauge scattering processes involving a low scale $W_R$ can rapidly wash out any primordial asymmetry generated above the mass scale of $W_R$. In fact, some of these gauge processes can continue to significantly reduce the generation of lepton asymmetry below the mass scale of $W_R$, thus the vector-like quark portals seem to be more promising in this case. Other alternatives include mechanisms like neutron-antineutron oscillation or some alternative LRSM scheme such as the Alternative Left-Right Symmetric Model \cite{Ma:1986we} where the dangerous gauge scatterings can be avoided by means of special gauge quantum number assignments of a heavy neutrino \cite{Dhuria:2015hta}. Also note that, in general, one can utilize the singlet neutral vector-like lepton as a dark matter candidate by ensuring the stability against decay into usual LRSM fermions. Finally, the real bi-triplet scalar field $\eta$ introduced to achieve successful gauge unification can also be a potential dark matter candidate.

In passing, we would like to mention that attempts have been made in the literature to address the broadness of the resonance using an invisible component of the scalar width. This in turn gives a large monojet signal which have been constrained from run-1 monojet searches at ATLAS \cite{Aad:2015zva} and CMS \cite{Khachatryan:2014rra}, see for example Ref. [39]. However, the monojet search data seems to disfavor the required rates to explain the broadness of the resonance. In our model, $S$ can couple to $XX^{\dagger}$ and $Y Y^{\dagger}$ etc. leading to decay of $S$ into them, which produces missing energy final state. This mode can be constrained from monojet searches as long as $M_{X,Y}<M_{S}/2$. Even without the scalar $S$ being directly coupled to $X$'s, its decay can produce a pair jets and $X$'s via the $\lambda_{XU}$ coupling term in Eq.~(21), which can again be constrained using dijet searches at ATLAS and CMS. In the discussion above we assume that these constrains are respected if $M_{X,Y}<M_{S}/2$. While for the case $M_{X,Y}>M_{S}/2$, the monojet and the dijet constraints are no longer applicable since in this case $S$ will decay via a loop of $X(Y)$'s.


\section{Conclusions}{\label{sec5}}
We have considered a unified framework to explain the recent diphoton excess reported by ATLAS and CMS around 750 GeV. The addition of vector-like fermions and a singlet scalar $S$ to LRSM but without a scalar bidoublet explains the fermion masses and mixing via a universal seesaw mechanism. The diphoton signal with $\sigma(pp \to S \to \gamma\gamma) \approx 4-12$~fb can be explained in this model with TeV scale vector fermions. The broad width suggested by the ATLAS excess cannot be understood, though.

We successfully embed this model within an SO(10) GUT framework by introducing a real bi-triplet scalar. This additional scalar, which contains a potential DM candidate, allows the gauge couplings to unify at the scale $10^{17.7}$~GeV. We also discuss further possibilities in this class of LRSM models with vector-like fermions for mechanisms of baryogenesis and DM.

\section*{ACKNOWLEDGMENTS}
FFD would like to thank Suchita Kulkarni for useful discussions. PP is grateful to the Department of Science and Technology (DST), Govt. of India INSPIRE Fellowship DST/INSPIRE Fellowship/2014/IF140299. The work of SP is partly supported by DST, India under the financial grant SB/S2/HEP-011/2013. The work US is supported partly by the JC Bose National Fellowship grant under DST, India.

\bibliographystyle{utcaps_mod}
\bibliography{diphoton_LHC}
\end{document}